\documentclass[a4paper,11pt]{article}
\usepackage{pos}
\usepackage{graphicx}
\usepackage[utf8]{inputenc}

\usepackage{lineno}
\title{Collaboration between high schools in Japan and Argentina for cosmic-ray research using CosmicWatches}
 \ShortTitle{Cosmic-ray}

\author*[a]{Takeshi Nakamori}
\author[b]{Mika Takanashi}
\author[c]{Rikako Kono}
\author[c]{Ryuta Saito}
\author[d]{Kyoka Ozaki}
\author[d]{Haruki Nakadai}
\author[c,e]{Kazuo S. Tanaka}
\author[f]{Ana Beatriz Prieto}
\author[f]{Marianela Pepe}
\author[f]{Juan Wehinger}
\author[f]{Lucio Martinez}
\author[c]{Chizuru Nose}
\author[g]{Haruhi Enomoto}
\author[c]{Ryosuke Kita}
\author[c]{Kyoka Maruta}

\affiliation[a]{Yamagata University, Department of Physics\\
 1-4-12 Kojirakawa, Yamagata, Japan}
\affiliation[b]{Yonezawa Kojokan High School\\
1101 Sasano, Yonezawa, Japan}
\affiliation[c]{Tohoku University, Department of Physics\\
6-3 Aramaki Aza-Aoba, Aoba-ku, Sendai, Japan}
\affiliation[d]{Tohougakkan High School\\
1-7-1 Chuou-minami, Higashine, Japan}
\affiliation[e]{Paul Scherrer Institut\\
Forschungsstrasse 111, 5232 Villigen PSI, Switzerland}
\affiliation[f]{Huechulafquen Science Club,
Jun\'{i}n de los Andes, Neuqu\'{e}n Province, Argentina}
\affiliation[g]{Tokyo Institute of Technology, Department of Earth and Planetary Science\\
2-12-1 Ookayama, Meguro-ku, Tokyo, Japan}

\emailAdd{nakamori@sci.kj.yamagata-u.ac.jp}

\abstract{
Cosmic rays are ubiquitous and readily available, making them a good teaching tool for particle and astrophysics by young students.
Tan-Q is an inclusive outreach and educational project, providing students in Japanese junior-high or high schools with research opportunities to join cosmic-ray and particle physics. In the Tan-Q framework, the students in each school conduct their research with help from mentors who are mainly undergraduate students. Researchers are also extensively involved through regular Zoom meetings and continuous communication on Slack. Some cases are inter-school, and some are international. This paper presents one of the Tan-Q activities of joint research between high schools in Japan and Argentina to observe cosmic-ray muons using CosmicWatches. Our primary goal is to investigate the muon flux differences due to the differences in circumstances like altitudes and geomagnetic field strengths. Those involved learn not only particle physics but also statistical data analysis methods.
}
\FullConference{37$^{\rm{th}}$ International Cosmic Ray Conference (ICRC 2021)\\		July 12th -- 23rd, 2021\\
		Online -- Berlin, Germany}

\begin{document}
\maketitle

\section{Introduction}
In recent years, the importance of learning through scientific research activities has been widely recognized in school science education in Japan, and such inquiry activities are being conducted in schools across the country. There are multiple perspectives, such as fostering logical thinking skills and active learning. The research themes and forms of activities vary from school to school, but they often depend on the school, teachers, and other factors. Moreover, there is sometimes a disparity between schools that have been selected for the Super Science High School Program and those that have not. There is no doubt that the field of particle and astrophysics plays a major part in modern physics. Even though Japan has several accelerators and observatories, it is difficult for high school students to choose their research theme. Some of the reasons for this may include the detectors and data acquisition systems required to measure particles and that it is not easy for high school students to use large facilities independently.
As well as the Japanese situation, Argentina is promoting Science, Technology, Engineering, and Mathematics (STEM) education through research-based learning, which has recently been incorporated into the official curriculum of the country. Moreover, the nation hosts one of the world's greatest cosmic-ray experiments, the Pierre Auger Observatory \citep{aug}.
Although such major facilities often promote educational programs in both countries, the opportunity is relatively limited for nationwide students with potential interests.

On the other hand, especially secondary muons, cosmic rays can be observed anywhere and easily studied by junior and high school students with proper detectors and guidance. Thus, cosmic rays are a good teaching tool for particle and astrophysics for young students. There have been some successful examples, such as (a part of) Quarknet \citep{qnet}. 
However, only a few cases have existed in Japan \citep{tera03}, 
 where researchers collaborate with junior and senior high school students to observe cosmic rays. This case was categorized as citizen science and was positioned as a collaborative project on a scientists' theme. In Japan, citizen science projects have been conducted in recent years, along with intelligent devices and network communication technologies. One recent and successful example involving particle measurements is the Thundercloud project \citep{yuasa} to observe gamma rays generated in thunderclouds. In these cases, scintillation detectors are provided by scientists and installed in schools for observation together with students.
 
Researches by junior and senior high school students can be divided into two major categories: research on their themes and so-called citizen science, as introduced above, where the researchers bring the objectives. Suppose a framework can comprehensively support both directions; it should be possible to provide opportunities for particle space research to a broader range of junior and senior high school students.

\section{Tan-Q framework}
\begin{figure}
\centering
\includegraphics[width=0.9\linewidth]{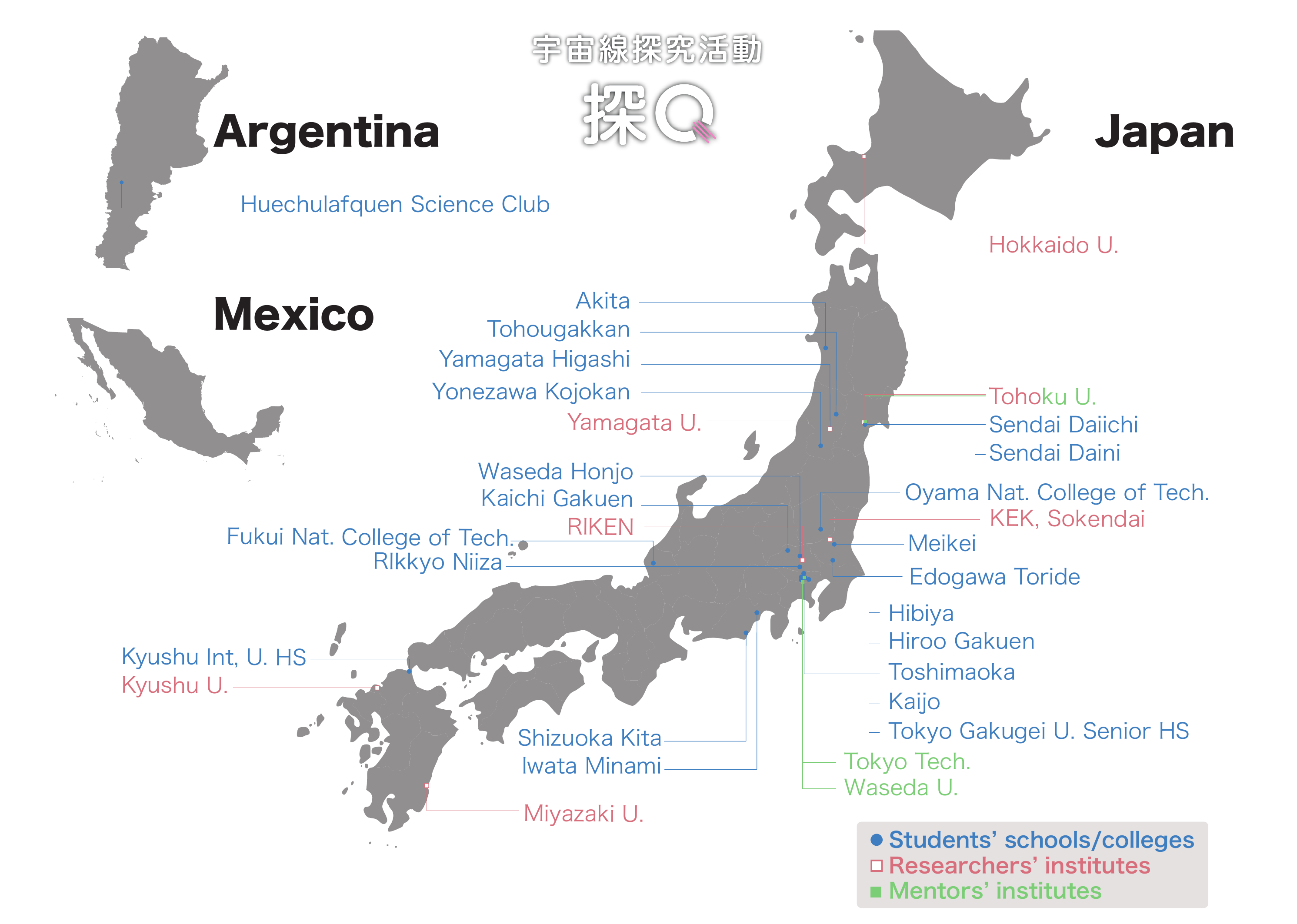}
\caption{Map showing Tan-Q's participating institutions. 
Circles indicate the locations of high schools and colleges of technology, 
and squares and triangles indicate those of the institutions to which the researchers and mentor students belong, respectively. }
\label{fig:map}
\end{figure}

Tan-Q
\footnote{Tan-Q's pronunciation is the same as the Japanese word corresponding to quest, seek and research}, 
inclusive outreach and educational project, was triggered along with such context and situation (PI: K. S. Tanaka, 2019--). Tan-Q consists of junior and senior high school students, mentor students from universities, and researchers. 
As shown in Figure~\ref{fig:map}, schools from nationwide Japan participate in the program and research a wide variety of topics. Some as individual schools and some as teams of students from different schools, including international, work on a single topic. We assign a dedicated mentor(s) or scientist(s) for each research group to
support the school students in their research activities. We contact each other through Slack daily and provide regular mentoring through Zoom to continuously support the students' research. In addition, we share files and meeting logs via Notion and Goole Drive. By making full use of these online tools, we realize sustainable activities without depending on the region.

Many research groups in Tan-Q use CosmicWatch (CW), a small and inexpensive desktop cosmic ray muon detector developed by \citet{cw} (see also section 3.2). 
Through the measurement workshops organized by Tan-Q, we make the CW known to the students and schools.
And then, the CW is distributed to schools that are interested in CW research activities. Some schools are using CW in its original form, for example, to search for a correlation between weather conditions and solar activity and cosmic ray rates. In other cases, they have modified the scintillator and data acquisition programs to research their topics.

In this paper, we present a case that is one of the Tan-Q activities with CWs, a collaboration between Japan and Argentina.

\section{Case: AR-JP joint research}
\subsection{Research objective}

The Earth's magnetic field has been monitored extensively on Earth's surface 
since the 1800s and also by satellites these days. 
There is an area known as the South Atlantic Anomaly (SAA), 
with a weak geomagnetic field strength around the South Atlantic Ocean. 
The SAA area is continuously growing, and its center is moving on a 100-year scale. 
This phenomenon reflects the dynamics of the Earth's interior and is a sign of a reversal of the Earth's magnetic field, 
which is being actively studied \citep{jav16}.
The geomagnetic field blocks cosmic rays (CRs) arriving at the Earth, and CRs with energies lower than the cutoff rigidity (typically a few GeV) cannot penetrate the magnetosphere. As a result, the magnetic field over the SAA is weaker than in other regions, and then the CRs, having a power-law spectrum, come into the atmosphere more frequently.

Since Argentina is close to the SAA, the arrival frequency of cosmic-ray muons is expected to be higher than in Japan. This study aims to detect the excess in the rate of Argentina due to the weak geomagnetic field strength. Since the rate of muons varies with altitude, it is necessary to correct this effect and then compare the results.
It should be noted that \citet{cec} performed cosmic ray measurements associated with SAA in Chile within the framework of QuarkNet.

\subsection{Students' activity : Measurements}

The CW \citep{cw} consists of a plastic scintillator of $\sim 5 \times 5 \times 1$\,cm$^3$ and a silicon photomultiplier.
Arduino Nano handles the data acquisition, which is triggered when the pulse height exceeds a threshold value. 
Then  pulse height, time stamp, and environmental data such as temperature are recorded for each event. 
PCB pattern and Arduino image files are available as open source. 
Our project provided the CWs assembled by the mentors to the observation teams in Japan and Argentina.
 They connected the CWs to a Windows PC via USB to read the data.

\begin{figure}[h]
\centering
\includegraphics[width=0.9\linewidth]{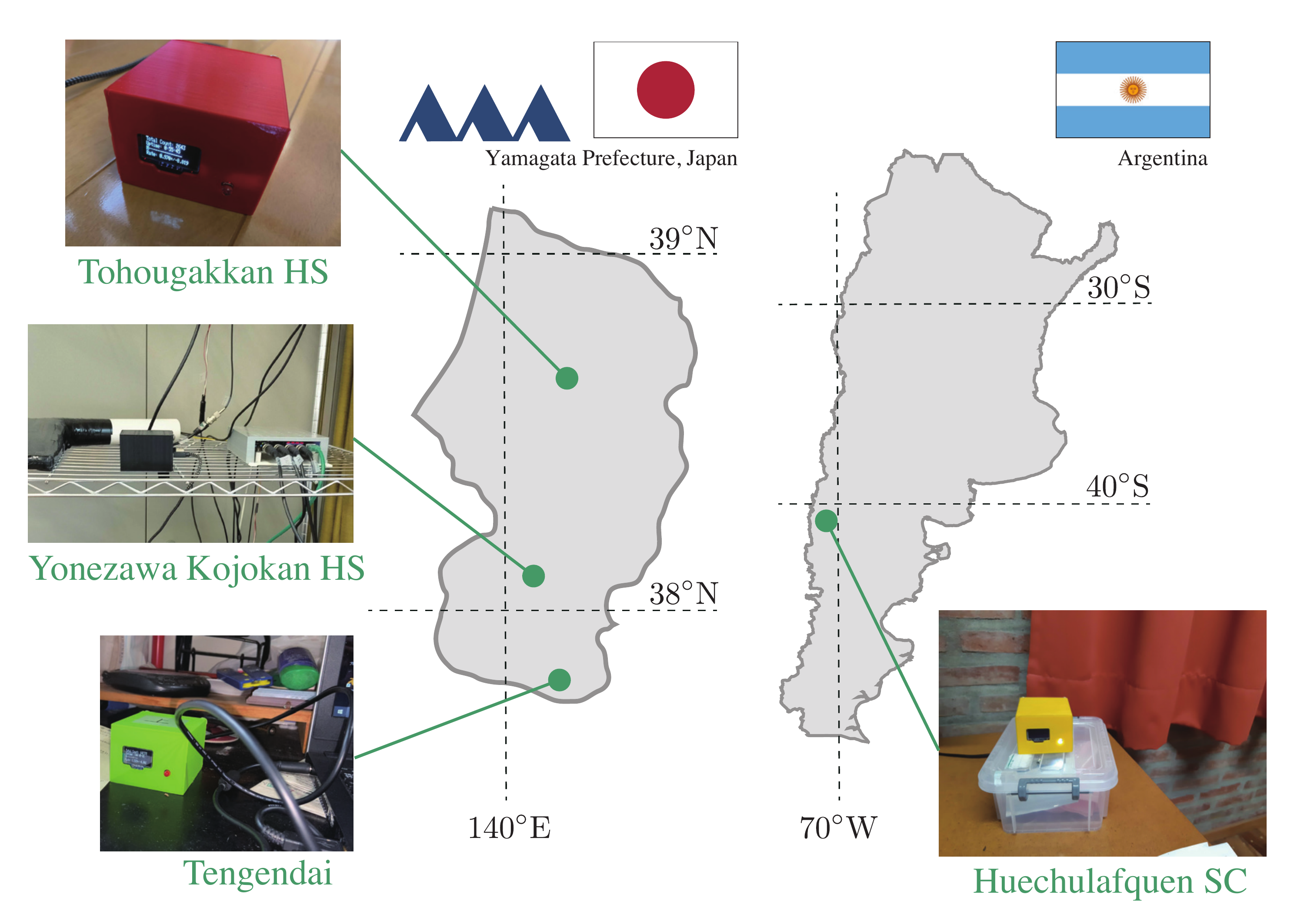}
\caption{Observation sites indicated in Yamagata prefecture, Japan (left), and Argentina (right). Note that these two maps are on a different scale.}
\label{fig:sites}
\end{figure}

Figure~\ref{fig:sites} shows the locations of the observation sites. We measured the average rate of muons at three sites in Yamagata prefecture, the northeastern district in Japan, to derive the correlation between altitude and frequency. Two sites of the three are high schools with relatively low altitudes, and the other is a kind of a pension at the ski resort with a high altitude above 1300\, m. By measuring within the minimal latitude and longitude range of Yamagata Prefecture, we can accurately track the differences in altitude.

In order to have the common solar condition which may affect the primary cosmic ray rate, we performed simultaneous measurements in Japan and Argentina during the same period. We took data continuously for several weeks from January to March of 2021, every 24 hours without interruption if possible. 
Then we shared the acquired data files on Google Drive both domestically and internationally. The mentors prepared such environments and the analysis template source codes on Google Colaboratory. Then the students analyzed the data with Python, mounting the Google Drive for efficient data handling.

\subsection{Students' activity : Analyses}
\begin{figure}[bth]
\centering
\includegraphics[width=0.95\linewidth]{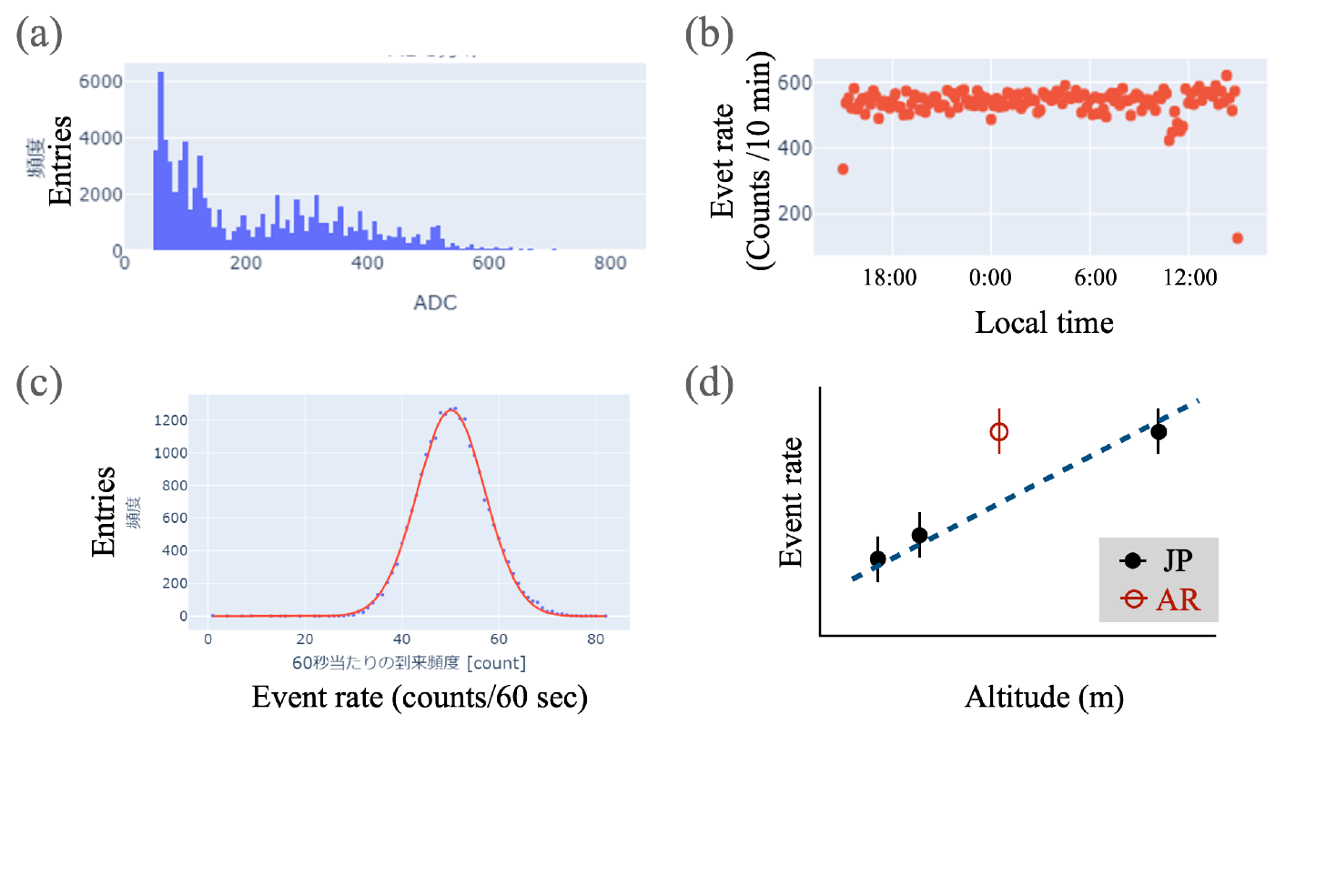}
\vspace{-1.5cm}
\caption{Summary of the analysis flow. 
(a) The ADC spectrum for a 1-day run. Two components are identified below/above the ADC value of 200. Apparent periodic spikes are artificial.
(b) Recorded event rate without the ADC cut. A slight drop can be found before 12:00, and such time intervals are removed from the analysis. 
(c) Histogram of the muon event rate derived from all-integrated data set of Argentina. With sufficient statistics, the distribution is well consistent with the Gaussian. 
(d) Sketch of the near future analysis (see text).}
\label{fig:active}
\end{figure}

The flow of the analysis is as follows. The first step is to apply a cut to the pulse-height ADC to select the qualified muon events. 
Figure~\ref{fig:active}(a) shows an example spectrum. 
We know that the event rate near the trigger threshold varies with temperature. 
If we choose the component around the peak at the higher ADC value, the rate stabilizes independent of temperature. 
Another team studied this finding in detail, working within the Tan-Q framework (Akita High School and Yamagata Higashi High School). 
It is also expected from the separate beta irradiation experiment that muons passing near the edge of the plastic scintillator also produce a low ADC component. In other words, muons striking near the edges are difficult to distinguish from thermal noise triggers. Since in this study, only one CW, or one scintillator, was used at each measurement location, we applied tighter cuts to the data aiming for accurate measurements of muons. 
In parallel, we cut the periods when data acquisition failed accidentally, mainly due to communication problems between the CW and the PC.
Figure~\ref{fig:active}(b) shows the event rates per 10 minutes in a time series. We exclude the periods with low rates deviating from the statistical fluctuations.
We applied these data cleaning procedures to all data sets for each location. 

Next, we obtained the count rate per minute from the surviving events and produced count rate histograms. Then we derived the average rate of muons by fitting with a Gaussian, as shown in Figure~\ref{fig:active}(c).
The rest of the project is in progress, and the students are currently learning statistics and error analysis. They are trying to fit a rate-altitude curve from data from three sites in Yamagata Prefecture, Japan, and test whether the rate in Argentina is significantly higher than that expected. 
Currently, we proceed to produce a figure like Figure~\ref{fig:active}(d).
The rate-alt dependence is expected to be exponential \citep{cwphys}, but since the altitude range is narrow, a linear fit could also be likely.

Further studies are worth consideration. In order to improve the accuracy of the rate-alt curve within Japan, it is conceivable to add data measured by other Tan-Q teams with CWs. Data from a very close area may be favorable, and the data collecting area is still limited to the relatively small region, Japan. The benefit of having more data points would be even significant. CW can connect two units and operate in coincidence mode \citep{cwphys}. This operation will significantly improve the signal-to-noise ratio of the muons at the hardware level. The more accurate the rate measurement will become, the more enhanced the sensitivity to the excess detection will be. Comparison of the muon rates with the models using EXPACS \citep{sato1, sato2} may be interesting.
Long-term monitoring, as long as a few years, could find an evolution of SAA. The CW can be a potent tool to see the global scale phenomena through easy desktop experiments.

\section{Concluding remark}
Although Tan-Q is still a young project, it functions on a wide scale in Japan with various activities supported by many mentors and researchers. Furthermore, through advanced research activities, which are sometimes beyond the standard curriculum, the project has proven
highly educational for junior and senior high school students and even mentor university students. 
From the students' point of view, Tan-Q promotes their research and, at the same time, allows them to see other junior and senior high school students engaged in similar research activities. They also connect with experts (sometimes outside Tan-Q) through regular debriefing sessions held twice a year. Most students who are conducting research activities in a single school cannot gain such experience. Furthermore, one of the proofs of the program's success is that some students went on to university through the activities of the Tan-Q and came back as mentors. In evaluating the results of the educational project, it is desirable to collect objective indicators
and data soon. Interviews are a possible method to manage the outcomes;
\begin{enumerate}
\item
How do junior and senior high school students change their interests in physics-related fields (not necessarily limited to particle and astrophysics)?
\item How the school teachers evaluate from their point of view?
\item What the mentors gain both as learners and educators through the Tan-Q activities?
\end{enumerate}

As for this Argentina-Japan project, it is rare both for Argentine and Japanese high school students to research with a contact outside the country. Therefore, this project is unique and noteworthy in the education in both countries, as it fosters an international perspective through the global-scale theme of studying the effects of geomagnetism.

\section*{Acknowledgment}
Tan-Q and this work is financially supported by Mitsubishi Memorial Foundation for Educational Excellence (Category 3, 2020).

%
%
%

\end{document}